# Grid Impact Analysis and Mitigation of En-Route Charging Stations for Heavy-Duty Electric Vehicles

Xiangqi Zhu, *Senior Member, IEEE*, Partha Mishra*, Member, IEEE,* Barry Mather, *Senior Member, IEEE*, Mingzhi Zhang*, Student Member, IEEE, and* Andrew Meintz

***Abstract*—This paper presents a consolidated grid impact analysis design and corresponding mitigation strategies for heavy-duty electric vehicle (EV) charging stations. The charging load of heavy-duty charging station can reach several megawatts, which could induce adverse impacts on the distribution grid if not effectively mitigated. To analyze the impacts and provide corresponding solutions, we select four representative distribution systems—including both single-feeder cases and a multi-feeder case—and design thorough test metrics for the impact analysis. The charging load profiles used in the analysis are derived from realistic conventional heavy-duty vehicle travel data. Based on the analysis results, charging stations are placed at three different representative locations in each distribution system: best, good, and worst locations. Mitigation strategies using a combination of smart charger functionality, on-site photovoltaic (PV) generation, and on-site energy storage (ES) are proposed and tested. A sizing method is also proposed to find the optimal PV-ES-charger capacity that minimizes the capital cost.**

## Nomenclature

| | |
|---|---|
| $a$ | Selected node for charging station placement |
| C | Total cost of the on-site PV-ES-charger ($) |
| $i$ | Node number in the system |
| $j$ | Node number in the system |
| $n$ | Total number of nodes in the system |
| $P_c^{max}$ | Maximum charging load (kW) |
| $P_{ref}$ | Maximum reactive power support needed (kvar) |
| $\|\delta P\|$ | A single-column matrix consisting of real power change on all nodes (kW) |
| $\delta P(j)$ | Real power change on node j (kW) |
| $\|p\|_{n\times 1}$ | A single-column matrix consisting of real power sensitivity factors related to node *a* |
| $p_{ij}$ | Sensitivity factor of real power |
| $Q_{charger}$ | Lowest reactive power capacity of charger (kvar) |
| $Q_{ref}$ | Maximum reactive power support needed (kvar) |
| $\|\delta Q\|$ | A single-column matrix consisting of reactive power change on all nodes (kvar) |
| $\delta Q(j)$ | Reactive power change on node j (kvar) |
| $q_{ij}$ | Sensitivity factor of reactive power |
| $S_{charger}$ | Charger capacity (kVA) |
| $S_{charger}^{set}$ | Settled charger capacity (kVA) |
| $S_{PV}$ | PV inverter capacity (kVA) |
| $\|VLSM_P\|$ | Voltage load sensitivity matrix for real power |
| $\|VLSM_Q\|$ | Voltage load sensitivity matrix for reactive power |
| $\|V\|$ | A single-column matrix consisting of voltages at all the nodes |
| $\|V_{ref}\|$ | A single-column matrix consisting of reference voltages for all the nodes |
| $\|V'\|$ | A single-column matrix consisting of the voltages for all the nodes after placing charging loads in the system |
| $\|\delta V\|$ | A single-column matrix consisting of the voltage Deviations calculated using $\|VLSM_P\|$ and $\|VLSM_Q\|$ |
| $\delta V(i)$ | Voltage deviation at node *i* |
| $\|\Delta V\|$ | A single-column matrix consisting of the voltage deviations after placing charging loads in the system |
| $\alpha$ | Coefficient between $\lambda_{ES-S}$ and $\lambda_{PV}$ |
| $\beta$ | Coefficient between $\lambda_{ES-P}$ and $\lambda_{PV}$ |
| $\lambda_{charger}$ | Unit cost of smart charger capacity ($/kVA) |
| $\lambda_{ES-P}$ | Unit cost of ES power capacity ($/kW) |
| $\lambda_{ES-E}$ | Unit cost of ES energy capacity ($/kWh) |
| $\lambda_{PV}$ | Unit cost of PV capacity ($/kVA) |

## I. Introduction

ELECTRIFICATION of trucks that are designed for regional freight transportation is expected to be realized in the near future as electric vehicle (EV) manufacturers are demonstrating these trucks on the road. In contrast to light-duty EVs such as sedans, sport utility vehicles, and pick-up trucks which used in daily life, the electric trucks are categorized as heavy-duty EVs that would require extreme high-power charging stations to provide much faster charging rates and would incur significantly large charging loads up to several megawatts on the distribution grid.

Therefore, it is critical to understand the grid impact that would be brought about by these heavy-duty EV charging

Xiangqi Zhu, Barry Mather, and Mingzhi Zhang are with the Power Systems Engineering Center. Partha Mishra and Andrew Meintz are with the Center for Integrated Mobility Sciences, all with the National Renewable Energy Laboratory, Golden, Colorado 80401, USA. (e-mails: xiangqi.zhu@nrel.gov, barry.mather@nrel.gov, mingzhi.zhang@nrel.gov, partha.mishra@nrel.gov, andrew.meintz@nrel.gov ).

stations for different locations on the grid and for different types of grids— and to develop appropriate mitigation solutions to ensure continued resilient grid operation.

However, in the state of the art, with respect to the grid impact brought by EV charging loads and the mitigation solutions, studies mainly focus on light-duty EVs [1]–[4]. Particularly, the emphasis of the provided mitigation solutions mostly lands on the understanding and coordination of the light-duty EV charging behavior for better grid operation [5]-[8]. The grid impact of the charging loads brought by heavy-duty EVs has not been thoroughly analyzed. In addition, there are no effective solutions in the literature to mitigate or resolve these impacts.

There are significant differences between the charging loads characteristics of the heavy-duty EV and the light-duty EV. To the distribution grid, charging loads of light-duty EVs are distributed, small loads that are comparable to individual household loads or small building loads. Therefore, the impact of those charging loads is dispersed across the grid and have a lower chance to induce concentrated serious impact.

In contrast, the loads of charging stations for heavy-duty EVs are heavily concentrated, significant high power spot loads, which are megawatt-level and comparable to the aggregated loads of hundreds or even thousands of households. To the distribution grid, those charging loads could induce large voltage sags and jeopardize system stability if no effective mitigation or management solutions can be implemented.

Because of the above important differences between the charging loads of light-duty EVs and heavy-duty EVs, the current EV grid impact analysis results in the literature cannot be used for the coming heavy-duty EVs.

Similarly, the existing grid impact mitigation solutions for light-duty EVs cannot be applied to the case of heavy-duty EVs. It is possible that system flexibility from other loads and distributed energy resources can be used to mitigate the impacts of light-duty EV charging loads [8]. Effective management and coordination of the charging behavior of the light-duty EVs can also be a solution to reduce the grid impact [5]-[7].

But for heavy-duty EVs, the system flexibility cannot help to a great extent because of the extreme loading of the charging station. Moreover, the charging management and coordination solutions for the light-duty EVs can't be applied to managing the heavy-duty EVs because of the fundamental difference of the functions between those two types of EV: the heavy-duty EVs are mostly used for freight transportation which usually have a planned routes and planned charging stops in the middle of the trip, while light-duty EVs are usually used for residents' daily life where the charging behavior can be managed by a number of different incentives such as charging price.

Furthermore, the management among different charging stations are not expected to provide enough help to the impact mitigation because a large power pulling will be induced as long as there is one heavy-duty EV charging in one station. Therefore, on-site solutions specifically designed for heavy-duty EV charging stations are more promising to mitigate the grid impact.

Hence, with the incoming heavy-duty EVs, it is urgent to conduct comprehensive grid impact analysis for the heavy duty EVs and develop effective impact mitigation solutions which can be used for the heavy-duty EV charging station.

To satisfy the above urgent need, in this work, we bridge the research gap and advance the state of the art in the following two aspects:

1) We develop a consolidated methodology for analyzing the grid impacts brought by heavy-duty EV charging stations. The analysis conducted using the developed methodology can provide a comprehensive understanding of the grid impacts and the grid hosting capacity for different sizes of heavy-duty EV charging stations at different locations on various types of grids.

2) We propose an effective on-site mitigation solution that can mitigate/resolve the voltage-related grid impacts with minimum capital cost.

The rest of the paper is organized as follows: Section II introduces the methodology of the grid impact analysis. Section III presents the proposed mitigation strategy. The case studies are presented in Section IV, and Section V concludes the paper and discusses future work.

## II. Grid Impact Analysis Methodology

This section introduces the proposed methodology for grid impact analysis, including test system preparation, system load modeling, charging load modeling, and test scenario design.

### *A. Test System Preparation*

We select four different distribution systems to create a representative portfolio of testing systems for the grid impact analysis and mitigation strategy examination. The IEEE 34-bus test system is selected to represent IEEE standardized systems; a realistic utility distribution system model is selected to represent a single-feeder case; a model of two connected utility feeders is selected to represent a multiple-feeder case; And a dedicated feeder that powers only the charging station without serving any other loads is derived from the aforementioned utility single distribution system. The selected distribution systems not only are representative for various system types, but also are suitable for heavy duty EV charging station placement. The IEEE 34 test system and the realistic single feeder are along the highway which provide locations for en-route charging stations. The two-feeder system is along a main road with a mix of residential and commercial loads and suitable for placing charging stations if EV truck is traveling in this area.

As shown in Fig. 1, the IEEE 34-bus system [9] has a long main line with several laterals. It has a total of 25 loads, including 6 spot loads and 19 distributed loads, comprising a total load of 1.8 MW during peak time.

Fig. 2 shows the model of a realistic utility distribution system. This distribution system has more than 2,500 nodes and more than 600 load nodes, including balanced three-phase loads and single-phase loads, with peak load reaching more than 5 MW.

As shown in Fig. 3, two connected realistic utility feeders are selected to represent the multiple-feeder distribution system case. The two feeders share one large substation transformer, with more than 3,500 nodes in total. The peak load of the two feeders together is approximately 6 MW.

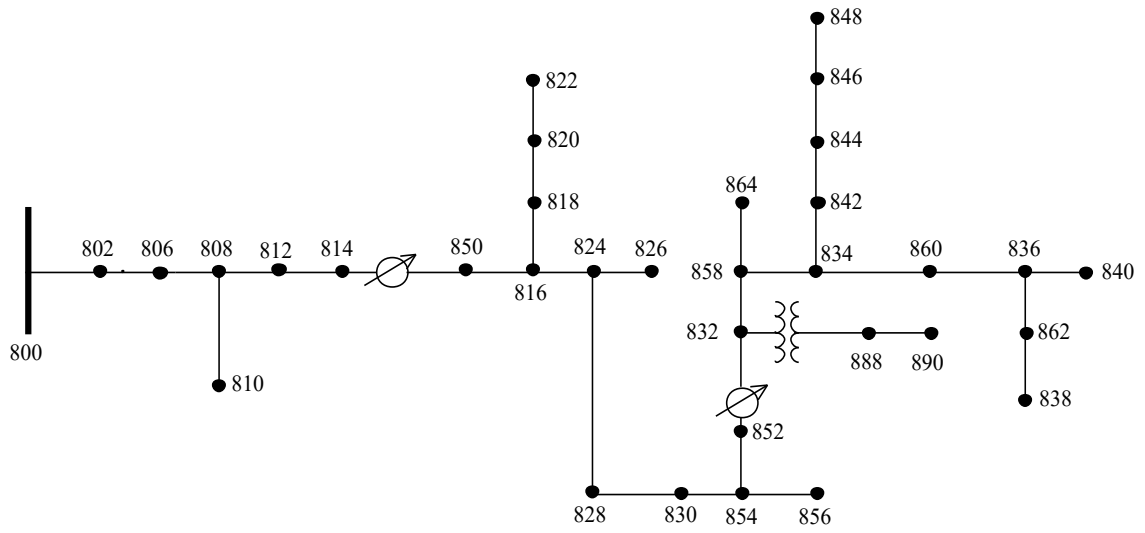

Fig. 1. Topology of IEEE 34-bus system

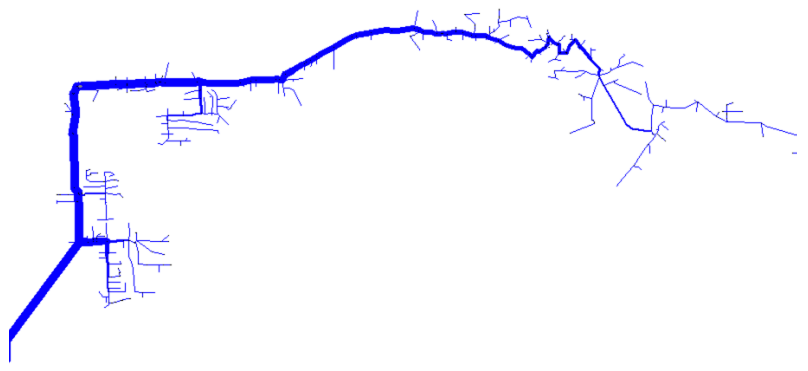
Fig. 2. Topology of the single-feeder case

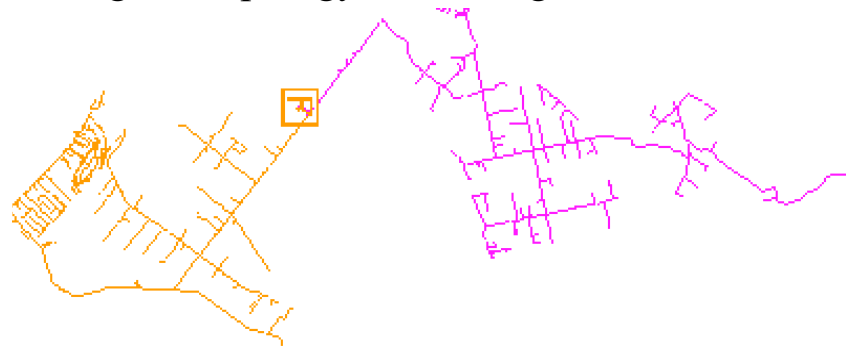
Fig. 3. Topology of the two connected feeders

By removing all the loads in the distribution system from the realistic single feeder case, we derive a dedicated feeder that serves only the charging station which serves as an example of a common utility practice for serving single large spot loads.

## B. System Load Modeling

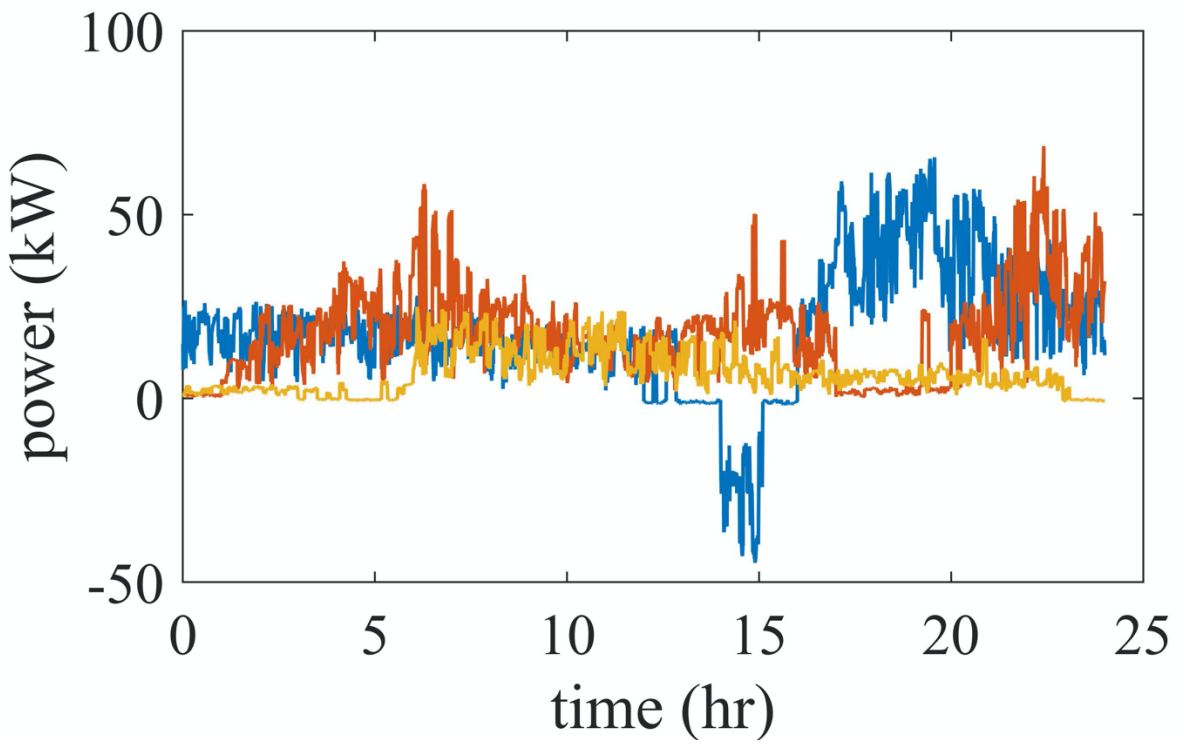

Fig. 4. Sample distribution nodal customer load profiles

By leveraging the synthetic load profile generation tool that was developed in [10]-[11], all the aforementioned distribution systems have been equipped with high-resolution, realistic load profiles. Instead of populating each bus with a load profile scaled from the substation load shape based on the transformer rating (i.e. standard load allocation), we use the diversity and variability libraries developed in [10]-[11] and generate diversified load profiles for the load buses. Thus, each load bus will have a unique load shape with appropriate variability. In this way, the high-resolution load profiles in each distribution system can have a realistic diversity factor, and each high-resolution load profile can have realistic variabilities. This allows us to better assess and analyze the grid impact under realistic loading conditions. Fig. 4 shows some sample load profiles generated from the load modeling tool for some of the customer nodes for the realistic single-feeder system.

## C. Charging Load Modeling

In this work, we use an in-house agent-based charging station modeling and analysis tool named Electric Vehicle Infrastructure, Energy Estimation, and Site Optimization (EVI-EnSite) tool to develop the station load profiles [12]–[13]. A flowchart of the vehicle charging operation in the EVI-EnSite tool and generation of station load profiles through a Monte Carlo simulation is shown in Fig. 5.

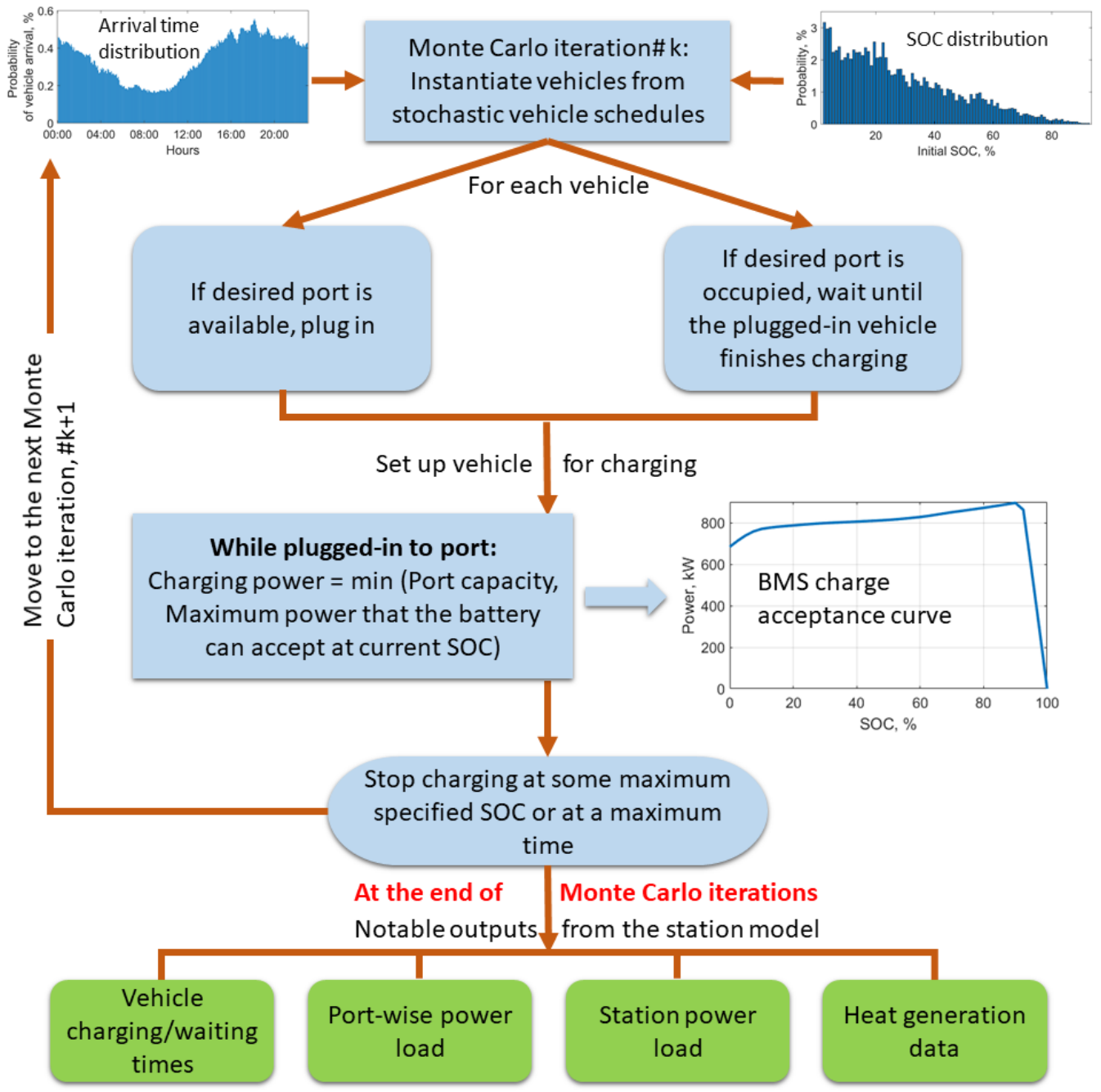

Fig. 5 Flowchart of a vehicle charging procedure in the EV-EnSite tool

The heavy-duty vehicles charging schedules are obtained by combining and analyzing real-world vehicle telemetry data analytics with EV system modeling [13]. Vehicle agents in this tool are defined using battery capacity, arrival time, initial state of charge (SOC), final desired SOC or energy demand, final stop time, and a charge acceptance curve. The charge acceptance curve is a proxy to emulate complex battery management system (BMS) control algorithms in a simplified manner for system-level simulations.

Similarly, a station agent is defined using the number of charging ports, power capacity of each port, and the station capacity. During simulation, a vehicle arriving at the station is either queued or plugged in depending on the availability of a desired port to charge. Charging continues until a maximum SOC is reached, or a desired amount of energy is added to the EV, or a stopping time criterion is met. We run the tool for several Monte Carlo iterations, each consisting of a station operational period of one month with a simulation time step of one minute. Section IV uses charging load profiles generated by EVI-EnSite for a vehicle traffic of 72 vehicles per day (with battery capacities ranging from 660-1200 kWh) and 1-port/3-ports/6-ports station configurations.

## D. Testing Scenario Design

To conduct a comprehensive analysis of the impact that the heavy-duty EV charging station could have on the grid, we design a series of testing scenarios as shown in Table I.

Table 1. Grid impact analysis test scenarios

| Test Scenarios | Station Location | | | Number of Port | | | Charging Load Pattern | | Feeder Load Pattern | | Total number of scenarios |
|---|---|---|---|---|---|---|---|---|---|---|---|
| | Best | Good | Worst | 1 | 3 | 6 | Daytime | Multi-shift | Residential | Commercial | |
| IEEE 34-bus system | X | X | X | X | X | X | X | X | X | X | 36 |
| Utility single feeder | X | X | X | X | X | X | X | X | X | | 18 |
| Utility multi-feeder | X | X | X | X | X | X | X | X | X | | 18 |
| Dedicated feeder | | | | | X | X | X | X | | | 4 |

Four aspects are considered:

1) *Charging station location.* Using the approach that we developed in [14], we rank the locations in the distribution system from best to worst. The best location will have the least voltage-related impact on the grid when the charging station is placed at this location, whereas the worst location will have the highest impact. Based on the ranking, we cluster the locations into three groups: best, good, and worst. Then we pick one representative location from each group and conduct the grid impact analysis. The details of ranking the locations on the grid can be found in [14]; generally, we leverage the voltage load sensitivity matrix developed in [15] and derive a voltage impact matrix to rank the locations.

2) *Number of charging ports at the charging station.* Three sizes of charging stations were considered: small, which has only one charging port; medium, which can serve 3 trucks at one time; and travel center, which can serve 6 trucks at the same time.

3) *Charging load pattern.* Two representative charging load patterns are considered here: daytime charging-dominated load, where the peak charging loads (megawatt-scale load) are concentrated during the daytime; and multishift charging load, where the peak charging load happens throughout the whole day (both day and night).

4) *System load pattern of the distribution system.* The charging load hosting capacity and the grid impact that the charging load might induce will be influenced by the original system load types because different load types have different characteristics (e.g., different peak times and valley times). Here, we select two representative load types to conduct the analysis: residential and commercial.

For the IEEE 34-bus test system, we design a comprehensive analysis that considers all the combinations among the four aspects, for a total of 36 unique scenarios. For the single-feeder and multi-feeder cases, we exclude the commercial load scenarios because the system loads on those two feeders are dominated by residential load. We evaluate 3-port and 6-port charging stations for the dedicated feeder case.

From the simulation results, we can summarize the maximum charging load and ramping rate each feeder can host on different locations without jeopardizing the grid and use that as a basis for the mitigation strategy development.

## III. Grid Impact Mitigation Strategy

This section introduces the mitigation strategies to mitigate/resolve the impacts to the grid brought by the heavy-duty EV charging stations.

### A. *PV-ES-Charger Solution*

The reactive power support is assumed as a function of the charger to effectively boost the voltage when the charging load is lowering the system voltage. However, when the charging load is significantly high, reactive power can no longer effectively raise the voltage, and some on-site real power generation is needed to offset part of the charging load. Therefore, an on-site PV system plus energy storage become a promising option.

As shown in Fig. 6 (a), when it is a cloudy day, we barely use the PV power in an effective manner because the PV power generation and the peak charging load is not temporally aligned. When we need a large amount of PV power to support the heavy charging load, the PV panel does not generate enough power. When it is a sunny day, we can effectively use the PV power to offset some demand from the charging station. However, there can also be an excess of PV power increasing the voltage during periods when the charging load is not high resulting in voltage above the upper limit, as shown by the blue dashed line in Fig. 6 (b). Note that in Fig.6, Vbase represents the voltage profile of the connection point of a 3-port charging station on two different good locations in the realistic utility single feeder system.

Therefore, it is important to design appropriate sizes for the three critical parts in the PV-ES-charger solution. In this way, the ES can effectively store excess PV power and power the charging load when PV is not generating, and the inverter can provide enough reactive power support when there is not enough on-site real power generation.

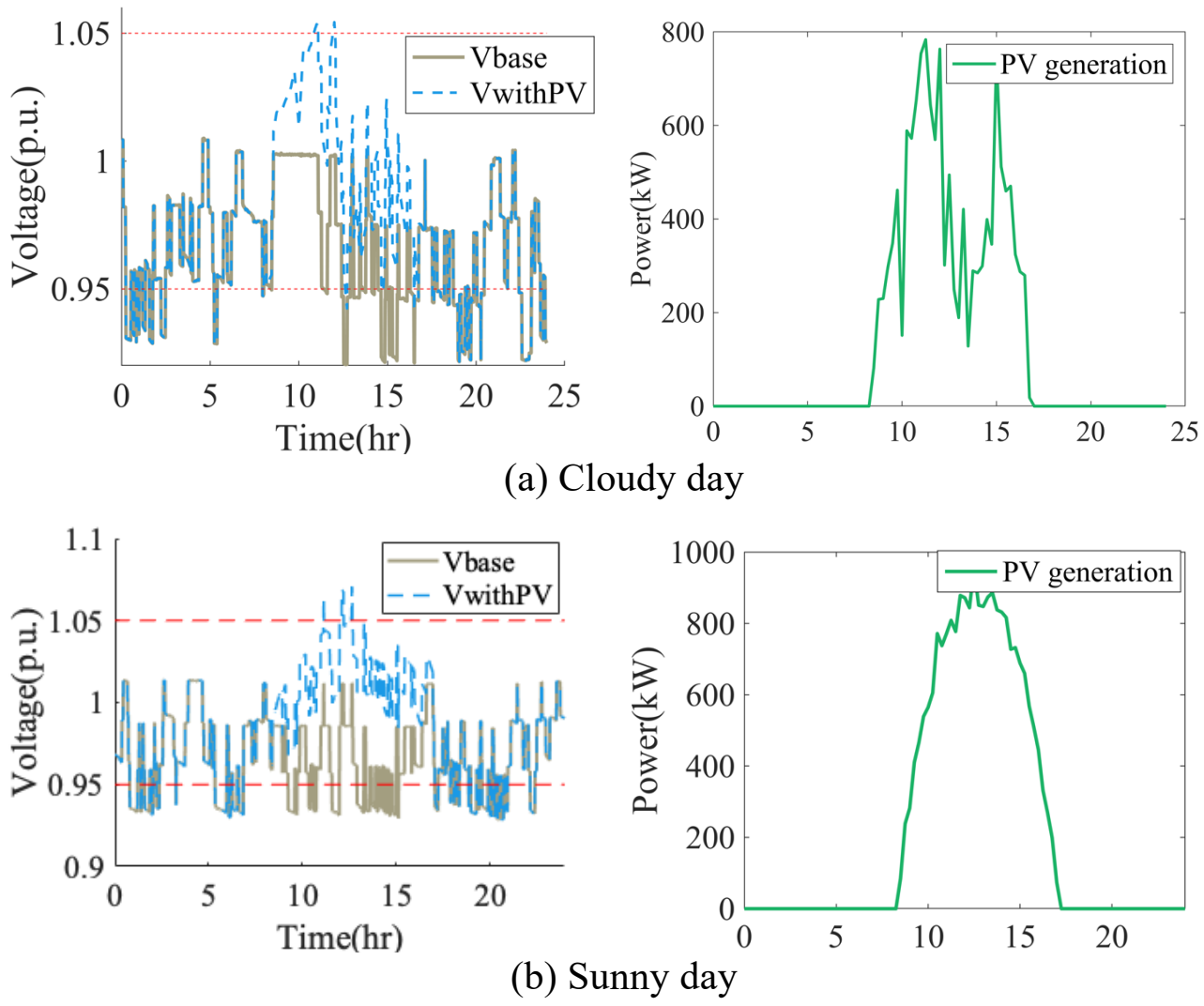

Fig. 6. Sample cases of voltage impact mitigation using on-site PV generation

### B. *PV-ES-Charger Sizing Strategy*

To balance the sizes of each part of the PV-ES-charger on-site solution, we propose a methodology to achieve the optimal size combinations of the PV, ES, and smart charger, with the objectives of maintaining the system voltage within limits and minimizing the total capital cost of the on-site PV-ES-charger solution.

As shown in (1–3), the voltage load sensitivity matrix (VLSM) developed in our previous work [15] can help calculate voltage changes in the system when the real/reactive power changes at one or multiple nodes in the system. We leverage the VLSM to estimate the voltages in the system when the peak charging load happens in the charging station, as shown in (4). The charging station is assumed to be placed at node $a$ in the system.

By calculating the difference between the system voltage at the peak charging load and the reference system voltage we would like to maintain, as shown in (6), we can calculate the maximum real/reactive power we need to maintain the system voltage at the reference level, as shown in (7) and (8). The $P_{ref}$ and $Q_{ref}$ then serve as the reference boundary for the following sizing coordination.

The maximum reactive power support the charger can provide at peak charging load is calculated in (9). The capacity of the PV needed is calculated in (10-11). Here $\eta$ decides how much reactive power the PV inverter can output when outputting maximum real power. Because reactive power cannot effectively boost the voltage when the load is consuming significant power. An amount of onsite generation needs to be guaranteed as shown in (12). The value of $\delta$ varies for different charging station scenarios.

As shown in (13), the total capital cost of the PV-ES-charger resolution can be calculated by plugging in the unit price of each part.

The main purpose of the ES is to store the extra PV energy and discharge to power the needed charging load. The size of the ES mostly depends on the interaction of the PV and the charging load profiles. We look at many different vehicle travel profiles coming to the station. The travel path of a particular vehicle in a portfolio might change, but the aggregated travel profile of this portfolio of vehicles are expected to be similar. Hence, a number of representative charging load profiles can demonstrate most charging situations at a station—therefore, for a specific charging station, the size of the ES depends on the PV generation profiles.

By analyzing the interaction of the typical PV generation profiles and charging load profiles, we can derive two coefficients—$\alpha$ and $\beta$—to represent the relationship between the ES size and PV size, as shown in (14). Then we can derive a cost equation that contains only two variables—namely, the size of charger and the size of PV—as shown in (15)–(17). A number of methods can be used to approximate the relationship between the ES and PV sizes, in this paper an empirical curve fitting method is used, where a number of PV and ES sizes ($E_{ES}$ and $P_{ES}$) are derived from representative PV and EV charging load profiles and then $\alpha$ and $\beta$ are approximated from the curve fitted from the empirical data points.

Substituting the $Q_{charger}$ in (10) with (9), and substituting the $S_{PV}$ in (16) with (11) , we can derive (18), which makes the cost the function of the $S_{charger}$.

The derivative of the cost function shown in (18) is formulated in (19). Eqs. (20) and (21) are derived from (19) and represent the situation when the derivative of the cost function is greater or less than zero, respectively.

We can observe five scenarios, as listed in Table 2. From scenarios 3–5, we can see that if the price of the charger is equal to or less than the price of the PV-ES, the total cost will always be reduced if we increase the size of the charger until we reach the maximum size needed. Scenario 1 shows that if the price of the charger is significantly higher than the price of the PV-ES, it is better to reduce the charger size as much as possible to minimize the total cost.

When the prices of the charger and PV-ES are comparable, we can have an optimal size of charger $S_{charger}^{set}$, which guarantees the minimized total cost, as shown in (22). In some cases where the price of charger is low, to guarantee the minimum size of onsite PV, the optimal charger size can't be reached, then the PV and charger sizes are calculated using (23)-(24). An example of the PV-ES-charger size selection will be presented in Section IV.

$$|\delta V| = |VLSM_P||\delta P| + |VLSM_Q||\delta Q| \quad (1)$$

i.e.,

$$\begin{vmatrix}\delta V(1)\\ \vdots \\ \delta V(n)\end{vmatrix} = \begin{vmatrix}p_{11} & \cdots & p_{1n}\\ \vdots & \ddots & \vdots\\ p_{n1} & \cdots & p_{nn}\end{vmatrix}\begin{vmatrix}\delta P(1)\\ \vdots \\ \delta P(n)\end{vmatrix} + \begin{vmatrix}q_{11} & \cdots & q_{1n}\\ \vdots & \ddots & \vdots\\ q_{n1} & \cdots & q_{nn}\end{vmatrix}\begin{vmatrix}\delta Q(1)\\ \vdots \\ \delta Q(n)\end{vmatrix} \quad (2)$$

We can derive from (2):

$$\delta V(i) = \sum_{j=1}^{n} p_{ij}\, \delta P(j) + \sum_{j=1}^{n} q_{ij}\, \delta Q(j) \quad (3)$$

$$|V'|_{n\times 1} = |V|_{n\times 1} - P_c^{max} \cdot |p|_{n\times 1} \quad (4)$$

$$|p|_{n\times 1} = \begin{vmatrix}p_{a1}\\ \vdots \\ p_{an}\end{vmatrix} \quad (5)$$

$$|\Delta V| = |V'| - |V_{ref}| \quad (6)$$

$$P_{ref} = -\sum_{i=1}^{n} {\Delta V(i)}/{p_{ai}} \quad (7)$$

$$Q_{ref} = -\sum_{i=1}^{n} {\Delta V(i)}/{q_{ai}} \quad (8)$$

$$Q_{charger} = \sqrt{{S_{charger}}^2 - P_c^{max\,2}} \quad (9)$$

$$Q_{PV} = (Q_{ref} - Q_{charger}) \quad (10)$$

$$S_{PV} = \frac{Q_{PV}}{\eta} \quad (11)$$

$$P_{PV} = S_{PV}\sqrt{(1-\eta^2)} > \delta P_{ref} \quad (12)$$

$$C = \lambda_{charger} S_{charger} + \lambda_{PV} S_{PV} + \lambda_{ES-E} E_{ES} + \lambda_{ES-P} P_{ES} \quad (13)$$

$$C = \lambda_{charger} S_{charger} + \lambda_{PV} S_{PV} + \lambda_{ES-E} \cdot \alpha S_{PV} + \lambda_{ES-P} \cdot \beta S_{PV} \quad (14)$$

$$C = \lambda_{charger} S_{charger} + (\lambda_{PV} + \alpha\lambda_{ES-E} + \beta\lambda_{ES-P}) \cdot S_{PV} \quad (15)$$

$$C = \lambda_{charger} S_{charger} + \lambda_{PV-ES} \cdot S_{PV} \quad (16)$$

Where:

$$\lambda_{PV-ES} = \lambda_{PV} + \alpha\lambda_{ES-E} + \beta\lambda_{ES-P} \quad (17)$$

$$C = f(S_{charger}) = \lambda_{charger} S_{charger} + \lambda_{PV-ES} \cdot \frac{1}{\eta}\left(Q_{ref} - \sqrt{S_{charger}{}^2 - P_c^{max\,2}}\right) \cdot {kVA}/{kvar}$$
(18)

Derivative of (18):

$$C' = \lambda_{charger} - 0.5\frac{1}{\eta}\lambda_{PV-ES}\frac{2S_{charger}}{\sqrt{S_{charger}{}^2 - P_c^{max\,2}}} \tag{19}$$

From (19), we can derive (20)–(21):

If $C' > 0$, then:

$$S_{charger} > \sqrt{\frac{\lambda_{charger}{}^2 P_c^{max\,2}}{\lambda_{charger}{}^2 - \lambda_{PV-ES}{}^2}} \tag{20}$$

If $C' < 0$, then:

$$S_{charger} < \sqrt{\frac{\lambda_{charger}{}^2 P_c^{max\,2}}{\lambda_{charger}{}^2 - \lambda_{PV-ES}{}^2}} \tag{21}$$

Table 2. Scenario overview

| Scenario Number | Condition | Description |
|---|---|---|
| 1 | $\lambda_{charger} \gg \lambda_{PV-ES}$ | C always increases as $S_{charger}$ increases |
| 2 | $\lambda_{charger} > \lambda_{PV-ES}$ | C always reduces when $S_{charger} < S_{charger}^{set}$ C always increases when $S_{charger} > S_{charger}^{set}$ |
| 3 | $\lambda_{charger} = \lambda_{PV-ES}$ | C always reduces as $S_{charger}$ increases |
| 4 | $\lambda_{charger} < \lambda_{PV-ES}$ | C always reduces as $S_{charger}$ increases |
| 5 | $\lambda_{charger} \ll \lambda_{PV-ES}$ | C always reduces as $S_{charger}$ increases |

Where: $$S_{charger}^{set} = \sqrt{\frac{\lambda_{charger}{}^2 P_c^{max\,2}}{\lambda_{charger}{}^2 - \lambda_{PV-ES}{}^2}} \tag{22}$$

$$S_{PV} = \frac{\delta P_{ref}}{\sqrt{(1-\eta^2)}} \tag{23}$$

$$S_{charger} = \sqrt{\left(Q_{ref} - \eta S_{PV}\right)^2 + P_c^{max\,2}} \tag{24}$$

## IV. Case Studies

This section presents representative grid impact analysis results from the simulations performed under the scenarios designed in Table 1 in Section II and presents a selected PV-ES-charger resolution. The grid simulations performed in this work are all conducted in OpenDSS with 1-minute resolution.

### *A. Representative Grid Impact Analysis Results*

We conduct two sets of simulations for all the defined scenarios shown in Table 1. For the first set, we place the charging station in the system without any mitigation strategies; whereas for the second set, we use a simplified power factor (PF) control method (PF:-0.9) to let the smart charger provide reactive power support to the grid connection point when the heavy-duty EVs are charging. As shown in Fig. 7, for each two bars of the same color, the left bar represents the set without mitigation resolutions, and the right bar represents the set with PF control as a mitigation method.

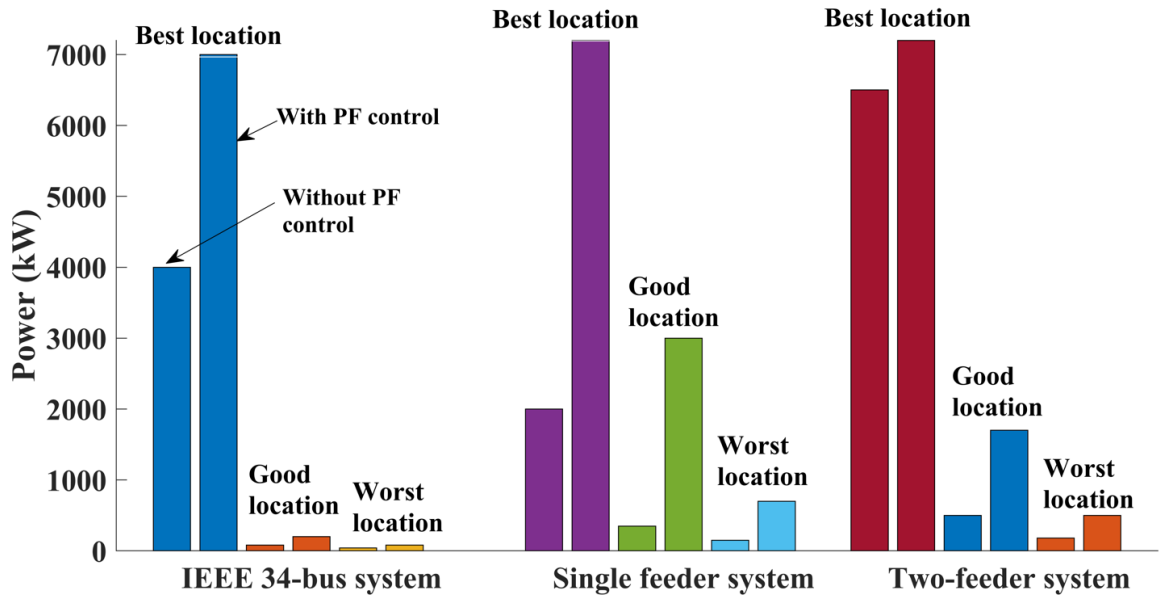

Fig. 7. Charging station hosting capacity analysis results

The maximum charging loads the system can handle for the selected best, good, and worst locations are derived from the time-series simulation results and summarized in Fig. 7. Note that the peak charging loads for the 1-port, 3-port, and 6-port charging stations are 1,200 kW, 3,600 kW, and 7,200 kW, respectively. The criteria of deciding if one location can host a size of charging load that all the voltages in the time-series simulations for this charging load profiles are within the limits ([0.95 p.u., 1.05 p.u.]).

As shown in Fig. 7, the three distribution systems can all host a 6-port charging station on the best location if we have a mitigation solution. The selected good and worst locations on the IEEE 34-bus system cannot host any charging station even with mitigation. The selected good location at the utility single-feeder site can host at maximum a 3-port charging station with mitigation, whereas the selected worst location can barely host a 1-port charging station. The selected good location on the two-feeder system can host a 1-port charging station, whereas the selected worst location cannot host charging stations. Although it is not shown in Fig. 7, the dedicated feeder can easily host a 6-port charging station.

Note that the results shown in Fig. 7 are based on the situation that the substation voltage is regulated at 1 p.u.; the feeders might be able to accommodate a larger charging station if the substation voltage is regulated at a higher level. Further, these analysis results are based on the simplified mitigation resolution that relies only on charger PF control to provide reactive power support. The impact should be further mitigated if on-site generation can be provided because there are low-voltage situations where real power can boost voltage, but reactive power effectively cannot.

Fig. 8 shows a 1-day time-series voltage profiles for Phase A of the grid connection point after placing a 3-port charging station on the different selected locations on the utility single feeder without any mitigation. The charging load pattern is multi-shift and the system is dominated by residential load.

We can see that the voltages are good and not heavily impacted by the charging loads at the best location, whereas the good locations have some voltage violations when heavy charging load occurs in the system. For the worst location, the voltages are significantly less than the lower limit representative of considerable adverse grid impact.

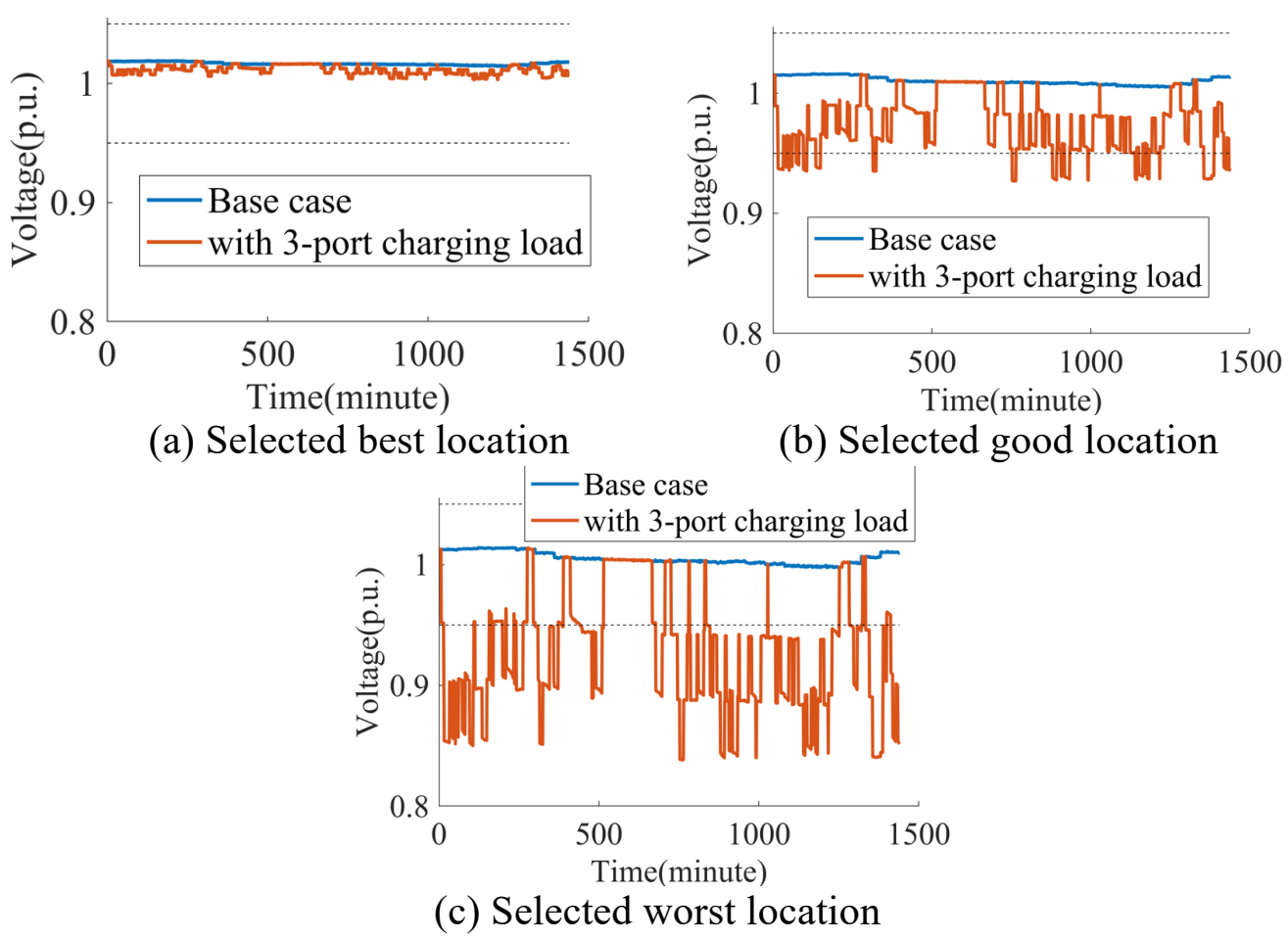

Fig. 8. One-day time-series simulation results with 3-port charging load on California single feeder

## B. PV-ES-Charger Resolution for Three-Port Station

To demonstrate the proposed PV-ES-charger resolution in Section III, we select the utility single distribution system to place a 3-port charging station on the selected good location in this system. The peak charging load is 3,600 kW, and the calculated $Q_{ref}$ is 1,294 kvar. The charging load pattern is multi-shift, and the system loads are residential loads.

The assumed prices of the charger and PV-ES system are shown in Table 3. These prices can be changed accordingly for different cases in real-world implementations. We select 7 days of representative charging load profiles and PV generation profiles to determine the ES size. Using the curve fitting method mentioned in Section III, $\alpha$ and $\beta$ in (14-15) are calculated as 4.75 and 1.

As heavy-duty EVs are mostly used for freight transportation, their travel routes are usually well-planned and not changing frequently. Therefore, we assume that 7 days' representative charging load profiles are a comprehensive set and cover most charging scenarios at a charging station. Also, the 7 days' PV profile we selected also constitute an inclusive set which cover 7 different weather conditions such as sunny, cloudy, rainy, etc.

For sensitivity analysis purpose, we've designed two price scenarios to analyze as examples (as shown in Table. 3). The PV-ES price is derived from the report in [16]-[17], while the charger prices are arbitrarily defined only for representative demonstration purpose. The price of charger in case 1 is comparable to the price of PV-ES while the price of charge in case 4 is four times of the price of the PV-ES. The cost curves of the two scenarios are shown in Fig. 9, with the corresponding PV/ES/charger sizes summarized in Table. 4. To guarantee an onsite PV size greater than $\eta P_{ref}$ , which is $20\% \times 1325 = 265\ kVA$, the selected charger size in case 1 is reduced from the optimal size and the PV capacity is increased to 265kVA. In case 2, the optimal charger size is in the practical range, which is used in the size finalization. The PV/ES/charger sizes derived in case 2 are used in the following analysis.

Table 3. Price and coefficients used in the case study

| Cases | $\lambda_{charger}$ ($/KVA) | $\lambda_{PV-ES}$ ($/KVA) | $\lambda_{ES-E}$ ($/KWh) | $\lambda_{ES-P}$ ($/KW) | $\lambda_{PV}$ ($/KW) |
|---|---|---|---|---|---|
| 1 | 5268 | 4489 | 661 | 350 | 1000 |
| 2 | 17956 | 4489 | 661 | 350 | 1000 |

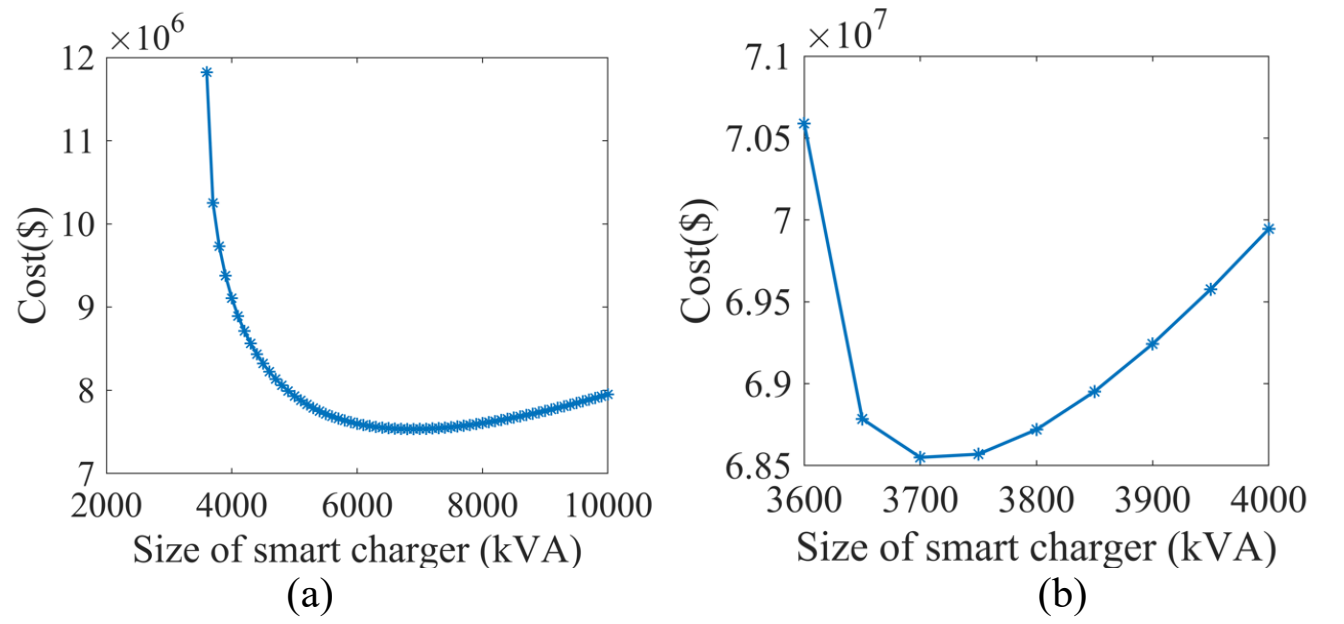

Fig. 9. Cost curves of two cases along with smart charger size

Table 4. PV-ES-charger size

| Cases | $S_{charger}^{set}$ (kVA) | $S_{PV}$ (kVA) | $E_{ES}$ (kWh) | $P_{ES}$ (kW) |
|---|---|---|---|---|
| 1 | 3752 | 265 | 1192 | 265 |
| 2 | 3718 | 364 | 1641 | 364 |

The PV generation profiles of 7 different days covering various weather conditions are shown in Fig. 10. Because of the different interactions between the charging load profiles and the PV generation profiles, the portion of the PV power effectively used to boost the voltage to mitigate the charging load impact is different for each day, as shown by the red lines in Fig. 10 and the red bars in Fig. 11.

Fig. 11 shows that a large portion of PV energy is not used to effectively mitigate the voltage impact if we let only the PV generate power and do not use the ES to store the extra PV power and discharge it when needed.

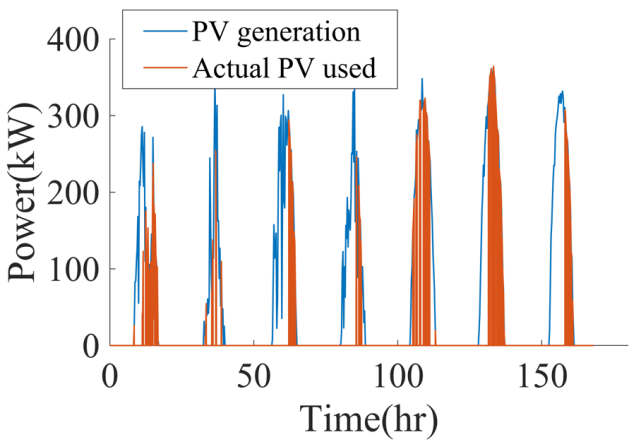

Fig. 10. PV generation profiles

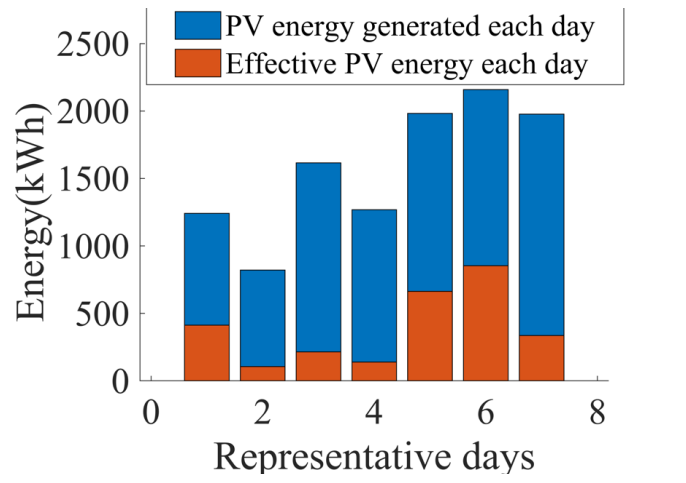

Fig. 11. Daily effective PV generation

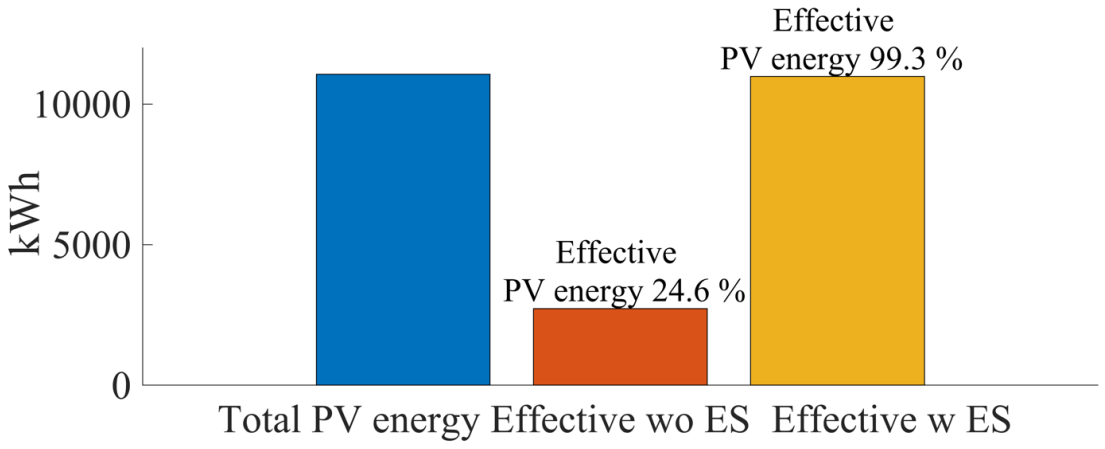

Fig. 12. Total effective PV generation

After we placed the ES (1641 kWh/364 kW) in the system, the total effective PV portion of the 7 days improved from 24.6% to 99.3%, as shown in Fig. 12. The ES saved almost 100% of the extra PV energy and enabled the on-site PV generation to effectively help mitigate the voltage impact.

Fig. 13 (a) shows the histogram of the voltages on all the nodes in the system in a 1-day simulation, whereas Fig. 13 (b) shows the voltage profile of the charging station connection point for the selected day, all under the no-mitigation scenario. Similarly, Fig. 14 (a) and (b) show the voltage distribution histogram and the voltage profile after placing the PV-ES-charger mitigation resolution in the system.

As shown, the mitigation solution successfully resolved the voltage impact brought about by the heavy-duty charging load and maintained the voltages within a reasonable range [0.95 1.05].

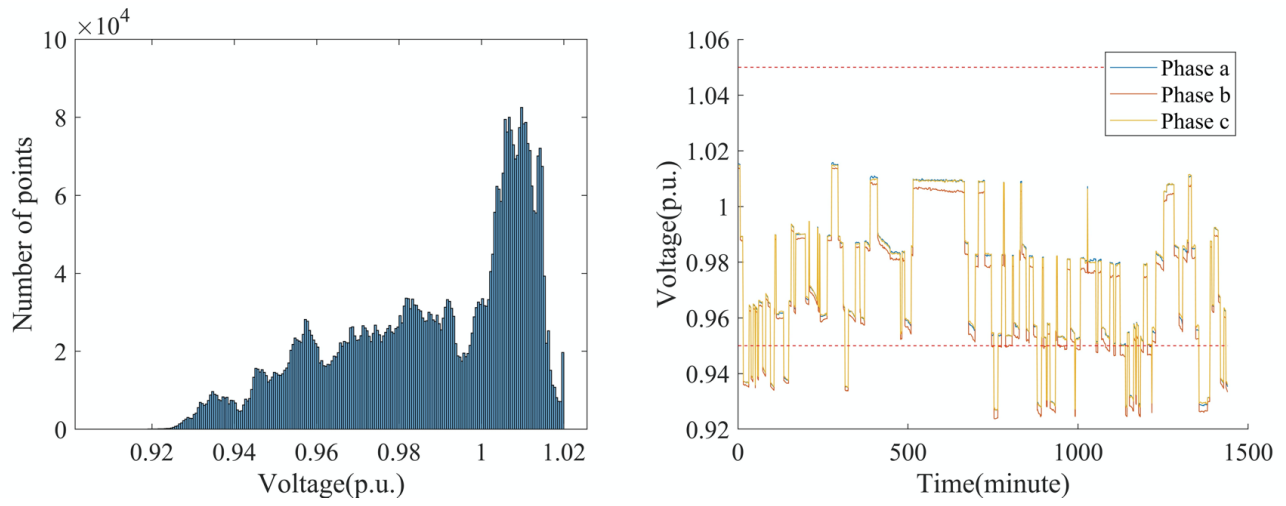

(a) Voltage distribution. (b) Voltage profile

Fig. 13. Voltage results without the mitigation resolution

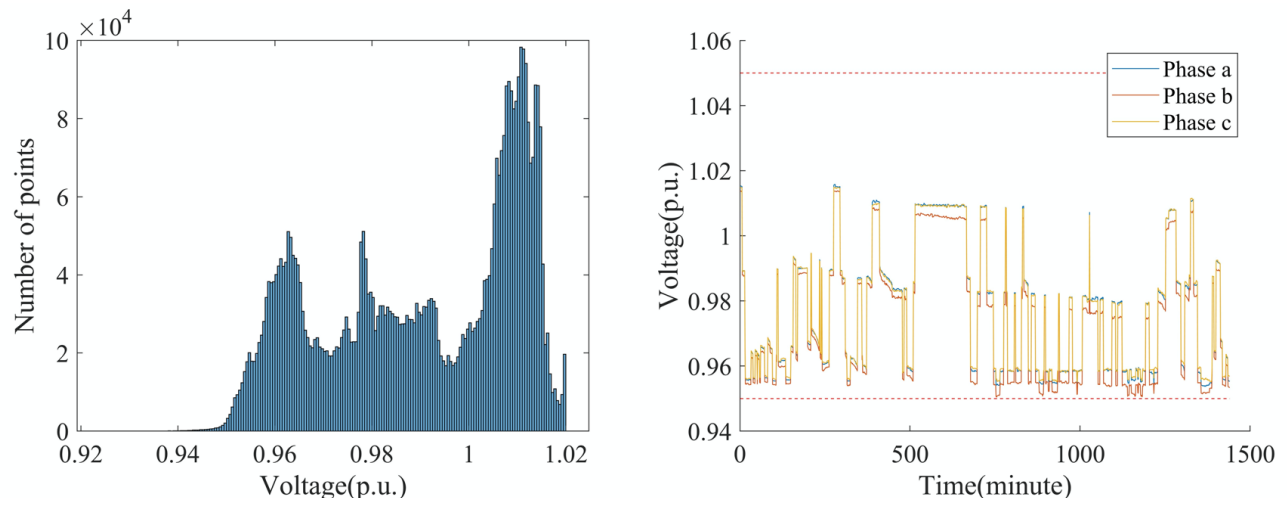

(a) Voltage distribution. (b) Voltage profile

Fig. 14. Voltage results with the mitigation resolution

## V. Conclusion and Future Work

In this work, we developed a consolidated analysis strategy for understanding the voltage impact that could be brought about by the integration of heavy-duty EV charging stations. Four representative distribution systems were selected for this analysis. Diversified scenarios that cover different combinations among charging station locations, sizes, and distribution system load patterns were designed. Furthermore, we proposed a PV-ES-charger on-site mitigation resolution to resolve the voltage impact with minimal capital cost.

The case study demonstrated that the proposed PV-ES-charger resolution can provide an optimal size combination of the PV, ES, and charger, which can minimize the cost of the resolution and help resolve adverse voltage impacts. Our next step is to incorporate charging load coordination among charging ports within the station to investigate whether we can further reduce the on-site PV and ES size with advanced charging load coordination and further investigate the optimal size of onsite PV/ES considering different cost scenarios.

Also, in our future work, we plan to analyze the impact on industrial feeder locations such as the impact brought by beginning point/destination parking and charging depot, and compare this impact with the impact on the commercial/residential feeders analyzed in this paper.

## Acknowledgments

This work was authored by the National Renewable Energy Laboratory, operated by Alliance for Sustainable Energy, LLC, for the U.S. Department of Energy (DOE) under Contract No. DE-AC36-08GO28308. Funding provided by U.S. Department of Energy Office of Energy Efficiency and Renewable Energy Vehicle Technologies Office via the 1+ MW Medium-Duty/Heavy-Duty Vehicle Project. The views expressed in the article do not necessarily represent the views of the DOE or the U.S. Government. The U.S. Government retains and the publisher, by accepting the article for publication, acknowledges that the U.S. Government retains a nonexclusive, paid-up, irrevocable, worldwide license to publish or reproduce the published form of this work, or allow others to do so, for U.S. Government purposes.